\documentclass{tPHM2e}
\usepackage{epsfig,amsmath,amssymb,txfonts,bm,hyperref}

\newcommand*{\et}[0]{\textit{et~al.}}
\DeclareMathOperator{\Tr}{Tr}

\newcommand*{\eps}[0]{\varepsilon}
\newcommand*{\ox}[0]{\otimes}
\newcommand*{\EE}[1]{\times 10^{{#1}}}

\newcommand*{\kB}[0]{k_{\text{B}}}

\newcommand*{\Tensor}[1]{\underline{#1}}
\newcommand*{\VECIT}[1]{\boldsymbol{\mathbf{#1}}}
\newcommand*{\TRANS}[1]{#1^\text{ts}}

\newcommand*{\Rv}[0]{\VECIT{R}}

\newcommand*{\xv}[0]{\VECIT{x}}
\newcommand*{\yv}[0]{\VECIT{y}}
\newcommand*{\zv}[0]{\VECIT{z}}
\newcommand*{\vv}[0]{\VECIT{v}}
\newcommand*{\uv}[0]{\VECIT{u}}

\newcommand*{\tv}[0]{\VECIT{t}}
\newcommand*{\dxv}[2]{\VECIT{\delta x}_{#1\to#2}}
\newcommand*{\bv}[1]{\VECIT{b}_{#1}}
\newcommand*{\dbv}[1]{\delta\bv{#1}}
\newcommand*{\domega}[0]{\delta\omega}
\newcommand*{\gv}[1]{\VECIT{\gamma}_{#1}}
\newcommand*{\ev}[0]{\VECIT{e}}
\newcommand*{\dbeta}[0]{\delta\beta}
\newcommand*{\dg}[0]{\delta g}

\newcommand*{\Eact}[0]{\Tensor{E}^\text{act}}
\newcommand*{\Vact}[0]{\Tensor{V}^\text{act}}
\newcommand*{\D}[0]{\Tensor{D}}
\newcommand*{\dd}[0]{\Tensor{d}}
\newcommand*{\ddc}[1]{d_{#1}}
\newcommand*{\prob}[1]{\rho_{#1}}
\newcommand*{\probt}[1]{\prob{\text{#1}}}
\newcommand*{\Et}[1]{E_\text{#1}}
\newcommand*{\dprob}[1]{\delta\prob{#1}}
\newcommand*{\preprob}[1]{\rho^0_{#1}}
\newcommand*{\trans}[2]{\lambda_{#1\to #2}}
\newcommand*{\transt}[2]{\trans{\text{#1}}{\text{#2}}}
\newcommand*{\dtrans}[2]{\delta\trans{#1}{#2}}

\newcommand*{\pretrans}[2]{\lambda^0_{#1#2}}
\newcommand*{\Etrans}[2]{\TRANS{E}_{#1#2}}
\newcommand*{\Etranst}[2]{\Etrans{\text{#1}}{\text{#2}}}
\newcommand*{\Eave}[0]{\langle E\rangle}
\newcommand*{\Ps}[1]{\Tensor{P}_{#1}}
\newcommand*{\Psc}[2]{P_{\text{#1},#2}}
\newcommand*{\Pst}[2]{P_{\text{#1},\text{#2}}}
\newcommand*{\PT}[2]{\TRANS{\Tensor{P}}_{#1#2}}
\newcommand*{\PTt}[2]{\PT{\text{#1}}{\text{#2}}} 
\newcommand*{\PTc}[3]{\TRANS{P}_{\text{#1}{\text{#2}},#3}} 
\newcommand*{\Pave}[0]{\langle \Tensor{P}\rangle}
\newcommand*{\deps}[0]{\delta\Tensor{\eps}}

\newcommand*{\R}[0]{R}
\newcommand*{\Rtv}[0]{\{\R,\tv\}}
\newcommand*{\Rtvinv}[0]{\{\R,\tv\}^{-1}}
\newcommand*{\Rtvinvexplicit}[0]{\{\R^{-1},-\R^{-1}\tv\}}

\newcommand*{\be}[0]{\begin{equation}}
\newcommand*{\ee}[0]{\end{equation}}
\newcommand*{\beu}[0]{\begin{equation*}}
\newcommand*{\eeu}[0]{\end{equation*}}
\newcommand*{\bme}[0]{\begin{multline}}
\newcommand*{\eme}[0]{\end{multline}}
\newcommand*{\bmeu}[0]{\begin{multline*}}
\newcommand*{\emeu}[0]{\end{multline*}}
\newcommand*{\ba}[0]{\begin{array}}
\newcommand*{\ea}[0]{\end{array}}
\newcommand*{\bfig}[0]{\begin{figure}[ht]}
\newcommand*{\efig}[0]{\end{figure}}
\newcommand*{\bfigwide}[0]{\begin{figure*}[t]}
\newcommand*{\efigwide}[0]{\end{figure*}}

\newlength{\wholefigwidth}
\newlength{\smallfigwidth}
\newlength{\halfsmallfigwidth}
\newcommand{\Fig}[1]{Fig.~\ref{fig:#1}}

\newcommand{\Eqn}[1]{Eqn.~\ref{eqn:#1}}

\newcommand*{\DOI}[1]{\href{http://dx.doi.org/#1}{doi://#1}}
\newcommand*{\arXiv}[1]{\href{http://arxiv.org/abs/#1}{arXiv:#1}}

\begin{document}

\title{Diffusivity and derivatives for interstitial solutes: Activation energy, volume, and elastodiffusion tensors}

\author{
\name{Dallas R. Trinkle$^{\ast}$\thanks{$^{\ast}$ Email: dtrinkle@illinois.edu}}
\affil{Department of Materials Science and Engineering, University of Illinois, Urbana-Champaign}
\received{\today}}

\maketitle
\begin{abstract}
Computational atomic-scale methods continue to provide new information about geometry, energetics, and transition states for interstitial elements in crystalline lattices. This data can be used to determine the diffusivity of interstitials by finding steady-state solutions to the master equation. In addition, atomic-scale computations can provide not just the site energy, but also the stress in the cell due to the introduction of the defect to compute the elastic dipole. We derive a general expression for the fully anistropic diffusivity tensor from site and transition state energies, and three derivatives of the diffusivity: the elastodiffusion tensor (derivative of diffusivity with respect to strain), the activation barrier tensor (logarithmic derivative of diffusivity with respect to inverse temperature) and activation volume tensor (logarithmic derivative of diffusivity with respect to pressure). Computation of these quantities takes advantage of crystalline symmetry, and we provide an open-source implementation of the algorithm. We provide analytic results for octahedral-tetrahedral networks in face-centered cubic, body-centered cubic, and hexagonal closed-packed lattices, and conclude with numerical results for C in Fe.
\end{abstract}


\maketitle

\section{Introduction}

Mass transport in solids is an integral part of material processing for materials from metals to ceramics to semiconductors. At the atomic scale, atoms move via defects: either vacancies for atoms on the crystalline lattice, or via interstitial sites off of the lattice\cite{Flynn1972,Allnatt1993}. For the case of interstitial defects, what had been often considered the ``simplest'' transport to model can often conceal surprising complexity\cite{Wu2011} when the interstitial network involves more than a single symmetry unique site. Furthermore, real materials also have non-homogeneous strain fields, and strain provides both a driving force\cite{Larche1982} for diffusion and modifies transport coefficients leading to a variety of complex behavior including: anisotropic dopant diffusion in semiconductor thin films\cite{Aziz1997,Daw2001,Aziz2006}, internal stress fields near a dislocation affecting transport and segregation of C solute atoms in Fe\cite{Veiga2010,Veiga2011} or vacancies in HCP metals\cite{Woo2000}, and anisotropy of dumbbell interstitial diffusion under biaxial stress in Cu and Pt\cite{Chan2008}. The influence of strain on diffusivity has been previously investigated only for cases where the interstitial random walk is purely uncorrelated\cite{Savino1977,Dederichs1978}; this excludes even the simple case of octahedral-tetrahedral networks in hexagonal-closed packed materials. While advances in density functional theory using modern supercomputers combined with transition-state finding methods\cite{Jonsson98,Henkelman2000} can compute the energies for different configurations and transition rates at the atomic scale with near chemical accuracy\cite{Janotti2004,Mantina2009}, general approaches to (a) automate the generation of a transition network, and (b) compute transport coefficients and related quantities using non-stochastic techniques are lacking.

The simplest transport coefficient to consider is the (generally anisotropic) diffusivity tensor $\D$, but we can also conceive of derivatives with respect to temperature and strain (or stress). An interstitial solute executes a random walk over a connected network of sites, where the transition rate between sites is given by a rate following an Arrhenius relation from transition-state theory\cite{Vineyard1957}. From the long-time behavior, the diffusivity tensor $\D$ can be computed, where $\D$ relates a steady-state flux $\VECIT{j}$ to a concentration gradient $-\nabla c$. The existence of different symmetry unrelated sites (and transitions) means that the total diffusivity may not have a simple Arrhenius temperature dependence given by a single activation barrier. Instead, over a sufficiently small temperature range, if we were to approximate the diffusivity tensor with an Arrhenius relation, $\D \approx \D_0 \exp(-\beta \Eact)$, then the negative derivative of diffusivity with respect to inverse temperature $\beta$ allows us to define
\be
\Eact := -\D^{-1/2}\frac{d\D}{d\beta}\D^{-1/2},
\label{eqn:ActivationEnergy}
\ee
corresponding to an anisotropic (and possibly temperature dependent) ``barrier'' tensor. The barrier should correspond to the activation energy for the rate-limiting transition in diffusion.
The fourth-rank elastodiffusion tensor\cite{Flynn1972,Dederichs1978},
\be
\dd := \frac{d\D}{d\Tensor{\eps}},
\label{eqn:ElastoDiffusion}
\ee
is the derivative of diffusivity with respect to strain. As both diffusivity and strain are symmetric tensors, $\ddc{abcd} = \ddc{bacd} = \ddc{abdc}$; furthermore, $\dd$ obeys  crystalline symmetry. Thus, it has a similar structure to elastic stiffnesses and compliances, with one exception: as it is not the second derivative of an energy, there is no general relationship between $\ddc{abcd}=dD_{ab}/d\eps_{cd}$ and $\ddc{cdab}=dD_{cd}/d\eps_{ab}$, unless imposed by crystalline symmetry.
Finally, activation volume can be defined in a similar fashion to activation energy, as a logarithmic derivative with respect to stress,
\be
\Vact_{abcd} := \kB T\,\D^{-1/2}\frac{d\D}{d\sigma}\D^{-1/2} = \kB T\sum_{ijkl=1}^3 \left(\D^{-1/2}\right)_{ai}\frac{d\D_{ij}}{d\eps_{kl}}\left(\D^{-1/2}\right)_{jb}S_{klcd},
\label{eqn:ActivationVolume}
\ee
and related back to the elastodiffusion tensor using the elastic compliances $S_{ijkl}$. This fourth-rank tensor contains the full aniostropy of diffusion response to stress. It can be reduced to a scalar activation volume by a double contraction,
\be
V^\text{act} := \frac13 \sum_{ij=1}^3 \Vact_{iijj}
\label{eqn:ScalarActivationVolume}
\ee
where the second contraction considers hydrostatic pressure, and the first contraction averages in all spatial directions.

In what follows, we develop the theory for the diffusivity, elastodiffusivity, activation barrier and volume tensors of an interstitial in terms of the site energies, transition state energies, and elastic dipoles (derivative of energy with respect to strain) of those same quantities. This includes the cases where the interstitial random walk includes correlation. We apply the theory to three simple examples: octahedral-tetrahedral networks in face-centered cubic, body-centered cubic, and hexagonal closed-packed lattices. We also provide numerical values for carbon diffusion in iron, using data from\cite{Veiga2011}. Finally, the work outlined here for the general case is implemented in open-source, publicly-available code (c.f.~Appendix~\ref{sec:implementation} and \cite{OnsagerCalc}) that can be applied to other systems\cite{Agarwal2016}.

\section{Methodology}

\subsection{Diffusivity}
In the case of interstitial diffusivity in the dilute limit, an interstitial atom moves through a network of sites in the crystal. Such a model allows simplification of the master equation: the system state is full described by the location of the interstitial, $\xv = \Rv + \uv$ for $\Rv$ a lattice vector and $\uv$ a vector in the unit cell. Due to translational invariance, the equilibrium site probability and set of transitions from a site are determined by the vector in the unit cell. For a site $i$ in the unit cell, it has an equilibrium site probability $\prob{i}$ that follows an Arrhenius relationship,
\be
\prob{i} := Z^{-1} \preprob{i} \exp\left(-\beta E_i\right)
\label{eqn:EquilProbability}
\ee
for site energy $E_i$, entropic prefactor $\preprob{i} = \exp(S_i/\kB)$ , and partition function $Z = \sum_i \preprob{i} \exp\left(-\beta E_i\right)$. The transition from site $i$ to site $j$ has a rate $\trans{i}{j}$,
\be
\trans{i}{j} := \frac{\pretrans{i}{j}}{\preprob{i}}\exp\left(-\beta\left[\Etrans{i}{j} - E_i\right]\right)
\label{eqn:TransRate}
\ee
for transition state energy $\Etrans{i}{j}$ and entropic prefactor $\pretrans{i}{j} = \exp(\TRANS{S}_{ij}/\kB)$, following \cite{Vineyard1957}. In this formulation, the transition state energy and entropic prefactors are equal for $i\to j$ and for $j\to i$, while it is not necessary that $\trans{i}{j}$ and $\trans{j}{i}$ are equal. Finally, the probabilities obey detailed balance, where $\prob{i}\trans{i}{j} = \prob{j}\trans{j}{i}$ for all $i,j$. Each transition results in a displacement of the diffusing atom by 
\be
\dxv{i}{j} := \xv_j - \xv_i.
\label{eqn:TransitionDisplacement}
\ee

With these definitions of site probabilities and rates, we can write down an expression for the diffusivity\cite{Allnatt1993}. There are two contributions to the diffusivity: an uncorrelated diffusivity and a correlation correction due to the unbalanced hops from individual sites. We define two rate matrices, $\Lambda_{ij}$,
\be
\Lambda_{ij} := \begin{cases}
\trans{i}{j} &: i\ne j\\
-\sum_j \trans{i}{j} &: i = j
\end{cases}
\label{eqn:LambdaMatrix}
\ee
which is the transition matrix for the master equation, and its symmetric counterpart
\be
\omega_{ij} := \prob{i}^{1/2} \Lambda_{ij} \prob{j}^{-1/2}.
\label{eqn:OmegaMatrix}
\ee
These matrices are negative-definite, and---due to detailed balance---both have a null-space related to the site probabilities $\prob{i}$. We define the scaled velocity vector (corresponding to the bias of jumps at a site as in \cite{Veiga2010,Veiga2011}),
\be
\bv{i} := \prob{i}^{1/2}\sum_j \trans{i}{j}\dxv{i}{j}
\label{eqn:VelocityVector}
\ee
which is non-zero when there are unbalanced hops from a site $i$; together with the bias-correction vector $\gv{i}$ which solves
\be
\sum_j \omega_{ij}\gv{j} = \bv{i}.
\label{eqn:GammaVector}
\ee
This is most easily done with the pseudoinverse of $\omega$, $g$, so that $\gv{i} = \sum_j g_{ij}\bv{j}$. Finally, the diffusivity is
\be
\D = \frac12 \sum_{ij} \dxv{i}{j}\ox\dxv{i}{j}\trans{i}{j}\prob{i} + \sum_i \bv{i}\ox\gv{i}
\label{eqn:Diffusivity}
\ee
where $\ox$ is the outer (or dyad) product of two vectors.%
\footnote{The construction $\VECIT{a}\ox\VECIT{b}$ is a second rank tensor such that $(\VECIT{a}\ox\VECIT{b})\cdot\vv = (\VECIT{b}\cdot\vv)\VECIT{a}$ for any vector $\vv$.}
Because $\omega$ is a symmetric matrix, so is $g$, and hence $\D$ is a symmetric second-rank tensor.

While the expressions leading up to \Eqn{Diffusivity} are general for non-interacting atoms hopping between sites, crystal symmetry can reduce the complexity of the expressions to be evaluated.  We use the Seitz notation\cite{Glazer2013} for a symmetry operation $\Rtv$, where for a point $\xv$, $\Rtv\xv:=R\xv + \tv$. Then, the inverse $\Rtvinv = \Rtvinvexplicit$. The full set of operations make up the \textit{space group}, and at any site $i$ there will be the subgroup of operations that leave that site fixed, its  \textit{point group}. If every site $i$ has a position $\xv_i$, then each group operation can be expressed as a matrix
\be
\Rtv_{ij} := \delta(\xv_i - \Rtv\xv_j)
\label{eqn:SpaceGroupMatrix}
\ee
where $\delta$ is the Kronecker delta. This matrix then applies the group operation $\Rtv$ to a site scalar $f_i$ as $\sum_j \Rtv_{ij} f_j$. A related matrix can be defined corresponding to vector at each site (such as $\bv{i}$ and $\gv{i}$), where with the orthonormal basis vectors $\{\ev_a\}$, then the matrix
\be
\Rtv_{ia,jb} := \Rtv_{ij}(\ev_a\cdot\R\cdot\ev_b)
\label{eqn:SpaceGroupMatrixVector}
\ee
transforms any site vector $\VECIT{f}_i$ as $\sum_a \ev_a\sum_{jb} \Rtv_{ia,jb}(\VECIT{f}_i\cdot\ev_b)$. All of our site scalars ($\prob{i}$), site vectors ($\bv{i}$ and $\gv{i}$), and site-to-site matrices ($\Lambda_{ij}$, $\omega_{ij}$, and $g_{ij}$) share the crystal symmetry, and so the following symmetry relations are true for any space group operation $\Rtv$,
\be
\begin{split}
\prob{i} &= \sum_j \Rtv_{ij} \prob{j},\\
\bv{i} &= \sum_a \ev_a \sum_{jb} \Rtv_{ia,jb} (\bv{j}\cdot\ev_b),\\
\gv{i} &= \sum_a \ev_a \sum_{jb} \Rtv_{ia,jb} (\gv{j}\cdot\ev_b),\\
\Lambda_{ij} &= \sum_{kl} \Rtv_{ik} \Lambda_{kl} \Rtv_{lj},\\
\omega_{ij} &= \sum_{kl} \Rtv_{ik} \omega_{kl} \Rtv_{lj},\\
g_{ij} &= \sum_{kl} \Rtv_{ik} g_{kl} \Rtv_{lj}.
\end{split}
\label{eqn:SymmetryRelations}
\ee
The implication of these symmetries is that $\prob{i}$, $\bv{i}$, and $\gv{i}$ can be written as linear combinations of vectors left unchanged by \textit{all} $\Rtv$ matrices, and that our matrices $\Lambda$, $\omega$, and $g$ can be entirely expanded in that same basis.

The basis functions for site scalars and site vectors come from the Wyckoff positions corresponding to sites in the network. Each Wyckoff position represents a full set of symmetry-related sites; any site scalar that obeys crystal symmetry will have the same value for each site corresponding to the same Wyckoff position. Hence, if we have a Wyckoff position that has $N_\text{W}$ sites in the unit cell, its basis vector components will be $N_\text{W}^{-1/2}$ for each site corresponding to that Wyckoff position, and 0 otherwise; there will be one basis function for each unique Wyckoff site \textit{regardless of how many sites that may represent in the unit cell}.
The Wyckoff positions also serve to construct a basis for site vectors combined with the point group operations for that site. Considering the point group operations that are available in a crystal, there are ten types: identity, a 2-, 3-, 4-, or 6-fold axis, mirror through a plane, or a 2-, 3-, 4-, or 6-fold axis combined with a mirror through that same plane. Note that a 2-fold axis combined with a mirror operation is inversion. As we are interested in vectors that remain \textit{unchanged} by all of the point group operations for a site, we can easily generate the basis corresponding to each operation. For identity, we have any 3-dimensional basis; for a mirror operation, a 2-dimensional basis that spans the mirror plane; for a 2-, 3-, 4-, or 6-fold axis, a 1-dimensional basis corresponding to the axis; and for a 2-, 3-, 4-, or 6-fold axis combined with a mirror, there are no vectors left unchanged. We construct the vector space left unchanged by each operation, and intersect to generate the final vector space for a single site. The spanning vectors for that site can then be rotated to the other sites corresponding to the same Wyckoff position, and normalized in a similar way to the site scalar basis.

The crystal symmetry divides diffusion networks into three types based on the crystal symmetry: networks without correlation, networks with correlation and inversion, and networks with correlation but without inversion. As it is possible for the site vector basis to be an empty set (e.g., if each Wyckoff position point group possesses a 2-, 3-, 4-, or 6-fold axis combined with a mirror operation), this excludes correlation \textit{explicitly by symmetry}. Next, there may be a non-empty site vector basis while the crystal has inversion. In this case, the transition matrices $\Lambda$ or $\omega$ expressed in the site vector basis are not singular. This is because the null-space vector of $\Lambda$---which corresponds to $\prob{i}$---has zero projection into any basis vector. Finally, in the most general case, the transition matrices remain singular, and so the construction of $g$ requires the pseudoinverse of $\omega$.

\subsection{Activation barrier}
We define the activation energy tensor from the first derivative of the diffusivity with respect to inverse temperature, and evaluate with perturbation theory. In this case, we introduce a small change in inverse temperature, $-\dbeta$, to $\prob{i}$ and $\trans{i}{j}$, then evaluate the first order change in $\D$. Then, the first order change in $\prob{i}$, $\dprob{i}$, is given by
\be
\begin{split}
\prob{i} + \dprob{i} &= \frac{\prob{i}e^{\dbeta\,E_i}}{\sum_j \prob{j}e^{\dbeta\,E_j}}\\
&= \prob{i} + \prob{i}\left(E_i -\Eave\right)\dbeta + O(\dbeta^2)
\end{split}
\label{eqn:DeltaProbEnergy}
\ee
where the average energy $\Eave = \sum_i \prob{i} E_i$. Next, the first order change in $\trans{i}{j}$, $\dtrans{i}{j}$, is given by
\be
\begin{split}
\trans{i}{j} + \dtrans{i}{j} &= \trans{i}{j}e^{\dbeta \left(\Etrans{i}{j} - E_i\right)}\\
&= \trans{i}{j} + \trans{i}{j}\left(\Etrans{i}{j} - E_i\right)\dbeta + O(\dbeta^2).
\end{split}
\label{eqn:DeltaTransEnergy}
\ee
From these, we can compute the related terms: $\dbv{i}$ and $\domega$. First,
\be
\begin{split}
\bv{i} + \dbv{i} &= (\prob{i} + \dprob{i})^{1/2}\sum_j (\trans{i}{j} + \dtrans{i}{j})\dxv{i}{j}\\
&= \bv{i} + \prob{i}^{1/2}\sum_j\trans{i}{j}\left[\Etrans{i}{j} 
-\frac12\left(E_i + \Eave\right)\right]\dxv{i}{j}\dbeta + O(\dbeta^2).
\end{split}
\label{eqn:DeltaBiasVectorEnergy}
\ee
Next for $\domega_{ii}$ we have
\be
\begin{split}
\domega_{ii} &= -\sum_j \dtrans{i}{j}\\
&= -\sum_j \trans{i}{j}\left(\Etrans{i}{j} - E_i\right)\dbeta + O(\dbeta^2)
\end{split}
\label{eqn:DeltaOmegaIIEnergy}
\ee
and for $\domega_{ij}$ ($i\ne j$) we have
\be
\begin{split}
\domega_{ij} &= (\prob{i}+\dprob{i})^{1/2}(\trans{i}{j} + \dtrans{i}{j})(\prob{j}+\dprob{j})^{-1/2}\\
&= \omega_{ij}\left[\Etrans{i}{j} - \frac12\left(E_i + E_j\right)\right]\dbeta + O(\dbeta^2).\\
\end{split}
\label{eqn:DeltaOmegaIJEnergy}
\ee
Finally, the first-order change in the pseudoinverse $g$, $\dg$ is given by (c.f.~Appendix~\ref{app:PseudoInverse})
\be
\begin{split}
g + \dg &= (\omega + \domega)^{-1}\\
&= g - g\,\domega\,g + O(\dbeta^2).
\end{split}
\label{eqn:DeltaGreenEnergy}
\ee
With all of these terms, we can compute the first-order correction to the diffusivity, which gives the derivative of $\D$ with respect to $-\beta$ by substituting into \Eqn{Diffusivity},
\be
\begin{split}
-\frac{d\D}{d\beta} &= \frac12 \sum_{ij} \dxv{i}{j}\ox\dxv{i}{j}\left[\Etrans{i}{j} - \Eave\right]\trans{i}{j}\prob{i} \\
&+ \sum_i \left(\frac{\dbv{i}}{\dbeta}\ox\gv{i} + \gv{i}\ox\frac{\dbv{i}}{\dbeta}\right) - \sum_{ij} \gv{i}\ox \frac{\domega_{ij}}{\dbeta}\gv{j}
\end{split}
\label{eqn:DeltaDiffusivityEnergy}
\ee
where the first term is the uncorrelated contribution and the remaining terms come from correlation. It should be noted that all of the perturbed site scalar, site vector, and matrices have the same symmetry as their unperturbed versions.  A similarly structured solution will follow for the evaluation of the elastodiffusion tensor.

\subsection{Elastodiffusion tensor}
To evaluate the elastodiffusion tensor, we introduce a perturbation through strain which changes site and transition state energies, as defined by the elastic dipole tensor. The elastic dipole tensor $\Ps{i}$ for a site $i$ is
\be
\Ps{i} := -\frac{dE_{i}}{d\Tensor{\eps}}.
\label{eqn:DipoleSite}
\ee
The elastic dipole can be conveniently evaluated in a supercell calculation from the stress in the cell: an interstitial is added to an initially undefected, unstressed supercell containing $N$ atoms (with equilibrium volume $V_0$ per atom), resulting in a stress $\Tensor{\sigma}$, then to first order in $N^{-1}$,
\be
\Ps{} \approx NV_0\Tensor{\sigma},
\label{eqn:DipoleSiteApprox}
\ee
which is straightforward to evaluate with density-functional theory methods; e.g., see \cite{Hanlumyuang2010,Varvenne2013,Garnier2014b,Kim2016}. Similarly, the energy of a transition state can also change with strain, as dictated by the elastic dipole tensor for the transition state $\PT{i}{j}$ for the transition state between $i$ and $j$,
\be
\PT{i}{j} := -\frac{d\Etrans{i}{j}}{d\Tensor{\eps}}.
\label{eqn:DipoleTrans}
\ee
This, too, can be approximated by the stress at the transition state in a supercell calculation as in \Eqn{DipoleSiteApprox}; e.g., see \cite{Veiga2011,Garnier2014b}. The definitions of elastic dipoles allow the introduction of a small strain perturbation $\deps$ to produce site energies changes $\delta E_i$ and transition energies $\delta\Etrans{i}{j}$ as
\be
\begin{split}
\delta E_i &= -\Ps{i}:\deps \\
\delta\Etrans{i}{j} &= -\PT{i}{j}:\deps
\end{split}
\label{eqn:DeltaEnergyStrain}
\ee
which is correct to first order in strain.%
\footnote{The double contraction sums over both indices of the two second-rank tensors: $A:B = \sum_{ab} A_{ab}B_{ba}$.}

We define the elastodiffusion tensor from the first derivative of the diffusivity with respect to strain, and evaluate with perturbation theory. Strain produces changes in site energies and transition state energies as laid out above, but also modifies all vectors in the network. The purely geometric contribution is evaluated by expanding the diffusivity tensor out in components as $\D = \sum_{ab} D_{ab} \ev_a\ox\ev_b$ for an orthonormal basis $\{\ev_a\}$; then a (symmetric) strain perturbs the basis vectors by $\delta\ev_a = \sum_c \delta\eps_{ac}\ev_c$, so that
\be
\delta\D^\text{(geom)} := \sum_{ab}\ev_a\ox\ev_b\left(\sum_c D_{ac}\delta\eps_{cb} + D_{bc}\delta\eps_{ca}\right)
\label{eqn:DiffusionGeometric}
\ee
which produces a (symmetrized) contribution to the elastodiffusion tensor
\be
d^\text{(geom)}_{abcd} := \frac12\left(\delta_{ad}D_{bc} + \delta_{ac}D_{bd} + \delta_{bc}D_{ad} + \delta_{bd}D_{ac}\right).
\label{eqn:ElastodiffusionGeometric}
\ee
To get the uncorrelated and correlated changes that we add to this, we get the first order correction to $\prob{i}$, $\dprob{i}$ as
\be
\begin{split}
\prob{i} + \dprob{i} &= \frac{\prob{i}e^{-\beta\delta E_i}}{\sum_j \prob{j}e^{-\beta\delta E_j}}\\
&= \prob{i} + \prob{i}\,\beta\left(\Ps{i} - \Pave\right):\deps + O(\deps^2)
\end{split}
\label{eqn:DeltaProbStrain}
\ee
where the average elastic dipole $\Pave = \sum_i \prob{i} \Ps{i}$. Next, the first order change in $\trans{i}{j}$, $\dtrans{i}{j}$, is given by
\be
\begin{split}
\trans{i}{j} + \dtrans{i}{j} &= \trans{i}{j}e^{-\beta \left(\delta\Etrans{i}{j} - \delta E_i\right)}\\
&= \trans{i}{j} + \trans{i}{j}\,\beta\left(\PT{i}{j} - \Ps{i}\right):\deps + O(\deps^2).
\end{split}
\label{eqn:DeltaTransStrain}
\ee
From these, we can compute the related terms: $\dbv{i}$---ignoring contributions of the strain to the hop vectors---and $\domega$. First,
\be
\begin{split}
\bv{i} + \dbv{i} &= (\prob{i} + \dprob{i})^{1/2}\sum_j (\trans{i}{j} + \dtrans{i}{j})\dxv{i}{j}\\
&= \bv{i} + \prob{i}^{1/2}\sum_j\dxv{i}{j}\trans{i}{j}\,\beta\left[\PT{i}{j} 
-\frac12\left(\Ps{i} + \Pave\right)\right]:\deps + O(\deps^2).
\end{split}
\label{eqn:DeltaBiasVectorStrain}
\ee
Next for $\domega_{ii}$ we have
\be
\begin{split}
\domega_{ii} &= -\sum_j \dtrans{i}{j}\\
&= -\sum_j \trans{i}{j}\,\beta\left(\PT{i}{j} - \Ps{i}\right):\deps + O(\deps^2)
\end{split}
\label{eqn:DeltaOmegaIIStrain}
\ee
and for $\domega_{ij}$ ($i\ne j$) we have
\be
\begin{split}
\domega_{ij} &= (\prob{i}+\dprob{i})^{1/2}(\trans{i}{j} + \dtrans{i}{j})(\prob{j}+\dprob{j})^{-1/2}\\
&= \omega_{ij}\,\beta\left[\PT{i}{j} - \frac12\left(\Ps{i} + \Ps{j}\right)\right]:\deps + O(\deps^2).\\
\end{split}
\label{eqn:DeltaOmegaIJStrain}
\ee
The expression for the first-order change in the pseudoinverse is the same as \Eqn{DeltaGreenEnergy}; hence, we can combine all of the contributions into \Eqn{Diffusivity} (similar to \Eqn{DeltaDiffusivityEnergy}) plus \Eqn{ElastodiffusionGeometric} to get
\be
\begin{split}
d_{abcd} &= \frac12\left(\delta_{ad}D_{bc} + \delta_{ac}D_{bd} + \delta_{bc}D_{ad} + \delta_{bd}D_{ac}\right) \\
&+ \frac12 \sum_{ij} \dxv{i}{j,a}\dxv{i}{j,b}\,\beta\left(\PT{i}{j,cd}-\Pave_{cd}\right)\trans{i}{j}\prob{i}\\
&+ \sum_i \left(\frac{\dbv{i,a}}{\delta\eps_{cd}}\gv{i,b} + \gv{i,a}\frac{\dbv{i,b}}{\delta\eps_{cd}}\right)
- \sum_{ij} \gv{i,a}\frac{\delta\omega_{ij}}{\delta\eps_{cd}}\gv{j,b}
\end{split}
\label{eqn:DeltaElastoDiffusion}
\ee
where the first term is the geometric contribution, the second term is the uncorrelated contribution, and the remaining terms come from correlation.

Introducing a finite strain into the lattice can reduce the symmetry of the lattice, and couples with the anisotropy of the elastic dipole tensors. The dipole tensors can be expanded in a site \textit{tensor} basis, similar to the site vector basis, using the point group for a single representative site of each Wyckoff position. We can also make a similar expansion for the transition states, by working with the ``double point group'': the set of symmetry operations that leave \textit{both} the initial and final positions unchanged. Despite the breaking of symmetry, we can see from \Eqn{DeltaElastoDiffusion}, that if a site $i$ has an empty vector basis (so that both $\bv{i}$ and $\gv{i}$ are zero by symmetry), the correlation contributions to the elastodiffusion tensor are \textit{also} zero. This means that the introduction of strain \textit{does not require} the computation of $\dbv{i}$ or $\delta\omega_{ij}$ for any sites where $\bv{i}$ and $\omega_{ij}$ was not already computed. However, it \textit{is} possible that particular values of $\deps$ can produce $\dbv{i}$ with \textit{components} that would be zero by symmetry for $\bv{i}$; the symmetry of $\dbv{i}/\delta\eps_{jk}$ is that of a third-rank tensor at site $i$.

\section{Results}
We can apply these results to three classic systems: octahedral-tetrahedral networks in face-centered cubic (FCC), body-centered cubic (BCC), and hexagonal closed-packed (HCP) structures. In the case of FCC and BCC, the network has zero site vectors basis (no correlation contribution), while the HCP network has a site vector basis oriented along the $c$-axis. In the case of BCC, tetrahedrals can be transition states, or states on their own. These three cases move from one, to two, to three different transition state energies. Finally, we conclude with a short set of numerical results for carbon in iron, using the EAM data generated by \cite{Veiga2011}.

\subsection{Face-centered cubic}
In a face-centered cubic lattice, the common interstitial sites are octahedral sites that can diffuse to tetrahedral sites, and vice versa. The face-centered cubic space group is $Fm\bar3m$, where the solvent atoms occupy $4a$ Wyckoff positions, with octahedrals at $4b$ and tetrahedrals at $8c$.\cite{Wyckoff} There is one octahedral site for every ``solvent'' atom, forming an interpenetrating FCC lattice. Each octahedral is connected to eight tetrahedral sites via $\frac{a}{4}\langle 111\rangle$ jumps. There are two tetrahedral sites for every lattice atom, forming an interpenetrating simple cubic lattice. Each tetrahedral is connected to four octahedral sites; the jump vectors are a subset of $\frac{a}{4}\langle 111\rangle$ vectors with tetrahedral symmetry (either $\{[111], [1\bar1\bar1], [\bar11\bar1], [\bar1\bar11]\}$ or  $\{[\bar1\bar1\bar1], [\bar111], [1\bar11], [11\bar1]\}$). By symmetry, there is a single transition state energy, $\Etranst{o}{t}$, so that we have two rates
\be
\begin{split}
\transt{o}{t} &\propto \exp\left[-\beta\left(\Etranst{o}{t} - \Et{o}\right)\right]\\
\transt{t}{o} &\propto \exp\left[-\beta\left(\Etranst{o}{t} - \Et{t}\right)\right]
\end{split}
\label{eqn:FCCrates}
\ee
and we have the probabilities
\be
\probt{o} = \frac{\transt{t}{o}}{2\transt{o}{t} + \transt{t}{o}},\quad
\probt{t} = \frac{\transt{o}{t}}{2\transt{o}{t} + \transt{t}{o}},
\label{eqn:FCCprob}
\ee
such that $\probt{o} + 2\probt{t} = 1$. The site vector basis for all of our sites are empty, so that we have no correlation terms to consider. From this, we have the isotropic diffusivity from \Eqn{Diffusivity},
\be
\begin{split}
\D &= \left\{8\frac12 \frac{a^2}{16}\transt{o}{t}\probt{o} +  4\cdot2\frac12 \frac{a^2}{16}\transt{t}{o}\probt{t}\right\}\mathbf{1}\\
&= \frac{a^2}{2}\transt{o}{t}\probt{o}\mathbf{1} = \frac{a^2}{2}\transt{t}{o}\probt{t}\mathbf{1},
\end{split}
\label{eqn:FCCdiffusivity}
\ee
and the isotropic activation energy from \Eqn{ActivationEnergy} and \Eqn{DeltaDiffusivityEnergy},
\be
\begin{split}
\Eact &= \D^{-1}\left\{8\frac12 \frac{a^2}{16}\left[\Etranst{o}{t}-\Eave\right]\transt{o}{t}\probt{o} +  4\cdot2\frac12 \frac{a^2}{16}\left[\Etranst{o}{t}-\Eave\right]\transt{t}{o}\probt{t}\right\}\mathbf{1}\\
&= \left[\Etranst{o}{t}-\Eave\right]\mathbf{1}.
\end{split}
\label{eqn:FCCactivationenergy}
\ee

We can also determine the elastodiffusion tensor and (isotropic) activation volume for an interstitial in FCC. To determine the elastodiffusion tensor, we note that the octahedral and tetrahedral elastic dipoles are fully isotropic by cubic and tetrahedral symmetry, respectively. The elastic dipole corresponding to the transition state has a three-fold rotation axis given by the $\langle 111\rangle$ vector connecting the two sites, so that the dipole has a component parallel $\PTc{o}{t}{\|}$ to the rotation axis $\vv = \frac{1}{\sqrt{3}}\langle 111\rangle$ and a component perpendicular $\PTc{o}{t}{\perp}$,
\be
\PTt{o}{t}(\vv) = \PTc{o}{t}{\|} \vv\ox\vv + \PTc{o}{t}{\perp}(\mathbf{1}-\vv\ox\vv) = \PTc{o}{t}{\perp}\mathbf{1} + (\PTc{o}{t}{\|}-\PTc{o}{t}{\perp})\vv\ox\vv
\label{eqn:FCCelasticdipole}
\ee
which, in component form, is
\be
\PTt{o}{t}([s_1,s_2,s_3])_{ab} = \PTc{o}{t}{\perp}\delta_{ab} + \left(\PTc{o}{t}{\|}-\PTc{o}{t}{\perp}\right)\frac{s_as_b}{3}
\label{eqn:FCCelasticdipolecomponents}
\ee
where $s_a = \pm1$ is the sign of the corresponding component of $\vv$. Similarly, $(\VECIT{\delta x}\ox\VECIT{\delta x})_{ab} = s_as_b a^4/16$. We can then evaluate the elastodiffusion tensor, using the Voigt notation for fourth-rank tensors, and \Eqn{DeltaElastoDiffusion},
\be
\begin{split}
\ddc{11} = d_{22} = d_{33} &= a^2\left\{ 1 + \frac\beta2\left(\frac23 \PTc{o}{t}{\perp} + \frac13\PTc{o}{t}{\|} - \Pave \right)\right\}\transt{o}{t}\probt{o}\\
\ddc{12} = d_{13} = d_{23} &= d_{21} = d_{31} = d_{32}\\
&= a^2\left\{ \frac\beta2\left(\frac23 \PTc{o}{t}{\perp} + \frac13\PTc{o}{t}{\|} - \Pave \right)\right\}\transt{o}{t}\probt{o}\\
\ddc{44} = d_{55} = d_{66} &= a^2\left\{ \frac12 + \frac\beta2\left(\frac13 \PTc{o}{t}{\|} - \frac13\PTc{o}{t}{\perp}\right)\right\}\transt{o}{t}\probt{o}
\end{split}
\label{eqn:FCCelastodiffusion}
\ee
and all other terms are zero, and where $\Pave = \Ps{\text{o}}\probt{o} + \Ps{\text{t}}2\probt{t}$. Due to symmetry, both $\Ps{\text{o}}$ and $\Ps{\text{t}}$ are isotropic (scalars). Then, the activation volume tensor from \Eqn{ActivationVolume} is 
\be
\begin{split}
  \Vact_{11} &= 2\kB TS_{11} + \frac13\left(\frac13\Tr\PTt{o}{t} - \Pave\right)B^{-1}\\
  \Vact_{12} &= 2\kB TS_{12} + \frac13\left(\frac13\Tr\PTt{o}{t} - \Pave\right)B^{-1}\\
  \Vact_{44} &= \frac12\kB TS_{44} + \frac16(\PTc{o}{t}{\|} - \PTc{o}{t}{\perp})S_{44}
\end{split}
\label{eqn:FCCactivationvolume}
\ee
for bulk modulus $B^{-1} = 3(S_{11}+2S_{12})$, and where $\Tr\PTt{o}{t} = 2\PTc{o}{t}{\perp} + \PTc{o}{t}{\|}$. The terms proportional to $\kB T$ all represent geometric contributions, while the remaining terms correspond to the sensitivity of the transition states and interstitial sites to stress.

\subsection{Body-centered cubic}
In a body-centered cubic crystal, the common interstitial site is an octahedral site, with the possibility of tetrahedral sites either as metastable or transition sites. Here, to compare with the face-centered cubic and hexagonal closed-packed case, we will take both octahedral and tetrahedral sites to be stable, and so there will be diffusion from octahedral to tetrahedral sites and between tetrahedral sites. The body-centered cubic space group is $Im\bar3m$, where the solvent atoms occupy $2a$ Wyckoff positions, with octahedrals at $6b$ and tetrahedrals at $12d$.\cite{Wyckoff} Unlike our other cases, there are three octahedral sites and six tetrahedral sites for every ``solvent'' atom; moreover, the octahedral sites do not have cubic point group symmetry, but rather tetragonal symmetry. If we consider a representative octahedral at $[00\frac12]$, it connects to four tetrahedrals in the $(001)$ plane, with jump vectors $\pm\frac{a}{4}[100]$ and $\pm\frac{a}{4}[010]$ and a transition state energy $\Etranst{o}{t}$. The elastic dipole associated with this octahedral has two components,
\beu
\Ps{\text{o}} = 
\Psc{o}{\perp}(\xv\ox\xv+\yv\ox\yv) + \Pst{o}{p}\zv\ox\zv,
\label{eqn:BCCoctelasticdipole}
\eeu
corresponding to the in-plane $\Pst{o}{p}$ and perpendicular $\Psc{o}{\perp}$ components. The transition state along $[100]$ has an elastic dipole with three components,
\beu
\PTt{o}{t} = 
\PTc{o}{t}{\|}\xv\ox\xv + \PTc{o}{t}{\perp}\yv\ox\yv + \PTc{o}{t}{\text{p}}\zv\ox\zv,
\label{eqn:BCCotelasticdipole}
\eeu
corresponding to the along-hop $\PTc{o}{t}{\|}$ as well as perpendicular and in-plane components. Considering next a representative tetrahedral at $[0\frac14\frac12]$, it connects to two octahedrals and four tetrahedrals, with jump vectors $\left\{\frac{a}{4}[01\bar1], \frac{a}{4}[011], \frac{a}{4}[\bar110], \frac{a}{4}[1\bar10]\right\}$ and a transition state energy $\TRANS{E}_\text{tt}$. The elastic dipole associated with this tetrahedral has two components,
\beu
\Ps{\text{t}} = 
\Psc{t}{\perp}(\xv\ox\xv+\zv\ox\zv) + \Pst{t}{a}\yv\ox\yv,
\label{eqn:BCCtetelasticdipole}
\eeu
corresponding to the four-fold rotation/mirror axis $\Pst{t}{a}$ and perpendicular $\Psc{t}{\perp}$ components. The transition state along $[01\bar1]$ has an elastic dipole with three components,
\beu
\PTt{t}{t} = 
\PTc{t}{t}{\|}\frac12\left(\yv-\zv\right)\ox\left(\yv-\zv\right) + \PTc{t}{t}{\perp}\frac12\left(\yv+\zv\right)\ox\left(\yv+\zv\right) + \PTc{t}{t}{\text{p}}\xv\ox\xv,
\label{eqn:BCCttelasticdipole}
\eeu
corresponding to the along-hop as well as perpendicular and in-plane components. In order for the site probabilities to match what was found in FCC, \Eqn{FCCprob}, we need to account for the increased multiplicity of octahedral and tetrahedral sites. We choose to keep the same normalization $\probt{o} + 2\probt{t} = 1$, and include a factor of 1/3 for each octahedral and tetrahedral site probability so that similarities to the face-centered cubic case can be most easily drawn.

Both the octahedral and tetrahedral sites have empty site vector bases, and so we do not need to consider correlation effects in the computation of the diffusivity or its derivatives. From this, we have the isotropic diffusivity of \Eqn{Diffusivity},
\be
\D = \frac{a^2}{12}\left\{\probt{o}\transt{o}{t} + 2\probt{t}\transt{t}{t}\right\}\mathbf{1},
\label{eqn:BCCdiffusivity}
\ee
and the isotropic activation energy from \Eqn{ActivationEnergy} and \Eqn{DeltaDiffusivityEnergy},
\be
\begin{split}
\Eact &= \D^{-1}\left\{\frac{a^2}{12}\left[\Etranst{o}{t}-\Eave\right]\probt{o}\transt{o}{t} +  \frac{a^2}{6}\left[\Etranst{t}{t}-\Eave\right]\probt{t}\transt{t}{t}\right\}\mathbf{1}\\
&= \frac{%
  \left[\Etranst{o}{t}-\Eave\right]\probt{o}\transt{o}{t} +  \left[\Etranst{t}{t}-\Eave\right]2\probt{t}\transt{t}{t}}{%
  \probt{o}\transt{o}{t} + 2\probt{t}\transt{t}{t}}
\mathbf{1}.
\end{split}
\label{eqn:BCCactivationenergy}
\ee
The activation energy changes from $\Etranst{o}{t}-\Eave$ when the octahedral to tetrahedral rate $\probt{o}\transt{o}{t}$ is the fastest to $\Etranst{t}{t}-\Eave$ when the tetrahedral to tetrahedral rate $\probt{t}\transt{t}{t}$ is the fastest. Finally, the elastodiffusion tensor has the cubic symmetry of the lattice, and using the Voigt notation for fourth-rank tensors with \Eqn{DeltaElastoDiffusion},
\be
\begin{split}
  \ddc{11} = \ddc{22} = \ddc{33} &= \frac{a^2}{12}\left\{ 2\left(\probt{o}\transt{o}{t} + 2\probt{t}\transt{t}{t}\right) + \frac\beta2\left(2\PTc{o}{t}{\|} - 2\Pave \right)\probt{o}\transt{o}{t} + \beta\left(\PTc{t}{t}{\|} + \PTc{t}{t}{\perp} - 2\Pave\right)\probt{t}\transt{t}{t}\right\}\\
  \ddc{12} = \ddc{13} = \ddc{23} &= \ddc{21} = \ddc{31} = \ddc{32}\\
&= \frac{a^2}{12}\left\{ \frac\beta2\left(\PTc{o}{t}{\perp} + \PTc{o}{t}{\text{p}} - 2\Pave\right)\probt{o}\transt{o}{t} + \beta\left(\frac12\PTc{t}{t}{\|} + \frac12\PTc{t}{t}{\perp} + \PTc{t}{t}{\text{p}} - 2\Pave\right)\probt{t}\transt{t}{t}\right\}\\
  \ddc{44} = \ddc{55} = \ddc{66} &= \frac{a^2}{12}\left\{ \left(\probt{o}\transt{o}{t} + 2\probt{t}\transt{t}{t}\right) + \beta\left(\frac12\PTc{t}{t}{\|} - \frac12\PTc{t}{t}{\perp}\right)\probt{t}\transt{t}{t}\right\}
\end{split}
\label{eqn:BCCelastodiffusion}
\ee
and all other terms are zero, and where
\be
\begin{split}
  \Pave &= \frac13\left\{\Tr \Ps{\text{o}}\probt{o} + \Tr \Ps{\text{t}}2\probt{t}\right\}\mathbf{1}\\
  &= \left\{\left(\frac23\Psc{o}{\perp} + \frac13\Psc{o}{\text{p}}\right)\probt{o} + \left(\frac23\Psc{t}{\perp} + \frac13\Psc{t}{\text{a}}\right)2\probt{t}\right\}\mathbf{1}.
\end{split}
\ee
Due to symmetry, $\Pave$ is isotropic even though $\Ps{\text{o}}$ and $\Ps{\text{t}}$ are not. Then, the activation volume tensor from \Eqn{ActivationVolume} is 
\be
\begin{split}
  \Vact_{11} &= 2\kB TS_{11} + \frac13\left\{\PTc{o}{t}{\|}S_{11} + \left(\PTc{o}{t}{\perp}+\PTc{o}{t}{\text{p}}\right)S_{12} - \Pave B^{-1} \right\}
  \frac{\probt{o}\transt{o}{t}}{\probt{o}\transt{o}{t} + 2\probt{t}\transt{t}{t}}\\
  &+ \left\{\frac12\left(\PTc{t}{t}{\|} + \PTc{t}{t}{\perp}\right)(S_{11} + S_{12}) + \PTc{t}{t}{\text{p}}S_{12} - \frac13\Pave B^{-1} \right\}
  \frac{2\probt{t}\transt{t}{t}}{\probt{o}\transt{o}{t} + 2\probt{t}\transt{t}{t}}\\
  \Vact_{12} &= 2\kB TS_{12}\\
  &+ \left\{\left(\PTc{o}{t}{\|} + \PTc{o}{t}{\perp} + \PTc{o}{t}{\text{p}}\right)S_{11} + \left(\PTc{o}{t}{\|} + \frac12(\PTc{o}{t}{\perp} + \PTc{o}{t}{\text{p}})\right)S_{12} - \frac13\Pave B^{-1} \right\}
  \frac{\probt{o}\transt{o}{t}}{\probt{o}\transt{o}{t} + 2\probt{t}\transt{t}{t}}\\
  &+ \left\{\left(\PTc{t}{t}{\|} + \PTc{t}{t}{\perp} + \PTc{t}{t}{\text{p}}\right)S_{11} + \left(\frac34\PTc{t}{t}{\|} + \frac34\PTc{t}{t}{\perp} + \frac12\PTc{t}{t}{\text{p}}\right)S_{12} - \frac13\Pave B^{-1} \right\}
  \frac{2\probt{t}\transt{t}{t}}{\probt{o}\transt{o}{t} + 2\probt{t}\transt{t}{t}}\\
  \Vact_{44} &= \frac12\kB TS_{44} + \frac18\left(\PTc{t}{t}{\|} - \PTc{t}{t}{\perp}\right)
  \frac{2\probt{t}\transt{t}{t}}{\probt{o}\transt{o}{t} + 2\probt{t}\transt{t}{t}}S_{44}
\end{split}
\label{eqn:BCCactivationvolume}
\ee
for bulk modulus $B^{-1} = 3(S_{11}+2S_{12})$. The terms proportional to $\kB T$ all represent geometric contributions, while the remaining terms correspond to the sensitivity of the transition states and interstitial sites to stress. There is more anisotropy in the activation volume for BCC compared with FCC, due to the lowered symmetry in the transition states.

\subsection{Hexagonal closed-packed}
In a hexagonal closed-packed crystal, there are a variety of interstitial sites available \cite{Wu2011,Middleburgh:2011dq,Zhang:2012fk,OHara2014,Agarwal2016}, but to compare with the our cubic cases, we will consider an octahedral-tetrahedral network. The hexagonal closed-packed space group is $P6_3mmc$, where the solvent atoms occupy $2c$ Wyckoff positions, with octahedrals at $2a$ and tetrahedrals at $4f$.\cite{Wyckoff} Similar to FCC, there is one octahedral and two tetrahedral sites for every ``solvent'' atom, but the connectivity is different. The octahedral sites form an interpenetrated simple hexagonal lattice, where each octahedral connects to two octahedrals directly above and below ($\pm\frac{c}{2}[0001]$ jumps) and six tetrahedrals (either $\{\pm\frac{a}{3}[1\bar10\frac{3c}{8a}], \pm\frac{a}{3}[01\bar1\frac{3c}{8a}], \pm\frac{a}{3}[\bar101\frac{3c}{8a}]\}$ or  $\{\pm\frac{a}{3}[\bar110\frac{3c}{8a}], \pm\frac{a}{3}[0\bar11\frac{3c}{8a}], \pm\frac{a}{3}[10\bar1\frac{3c}{8a}]\}$).%
\footnote{The tetrahedral sites are not required to sit at a $c/8$ distance away from the basal plane---only that their positions be symmetric under a mirror through the basal plane. Hence, the general case would introduce the Wyckoff parameter $z$ to character this position in the $4f$ site. However, in the case of diffusion, only the long-range contribution matters, so that change in jump vectors due to $z$ exactly cancels out, and we use $z=1/8$ for simplicity.}
The tetrahedrals are each connected to one other tetrahedral (either $\frac{c}{8}[0001]$ or $\frac{c}{8}[000\bar1]$) and three tetrahedrals (opposing the octahedral jumps, and with the $c$-axis contribution negating the tetrahedral jump contribution). This introduces three transition state energies, $\Etranst{o}{t}$, $\Etranst{t}{t}$, and $\Etranst{o}{o}$ with corresponding rates among octahedrals and tetrahedrals. The site probabilities match what was found in FCC, \Eqn{FCCprob}. From this, we can generate the anisotropic diffusivity tensor (take $\xv$ and $\yv$ to be orthonormal basis vectors in the basal plane and $\zv$ along the $c$-axis),
\be
\begin{split}
\D &= 
\frac{c^2}{4}\zv\ox\zv \probt{o}\transt{o}{o}
+ 2\left(\frac32 a^2\xv\ox\xv + \frac32 a^2 \yv\ox\yv\right) \probt{o}\transt{o}{t}
+ 3\cdot2\left(\frac{c^2}{64}\zv\ox\zv\right) \probt{o}\transt{o}{t}
+ \frac{c^2}{16}\zv\ox\zv \probt{t}\transt{t}{t} + \bv{\text{t}}\ox\gv{\text{t}}\\
\end{split}
\label{eqn:HCPDiffusivityUncorrelated}
\ee
once we solve for the velocity and bias-correction vectors. The octahedral sites possess a threefold axis (pointing along $\zv$) and a mirror plane perpendicular to that axis, so there is no vector basis at those sites; the tetrahedral sites also possess the same threefold axis, but no mirror plane, and so a vector can point along $\zv$ at the tetrahedral sites. A mirror operation through the basal plane maps one tetrahedral sites to the other connected by a tetrahedral-to-tetrahedral jump; thus, the single site vector basis alternates $\pm\zv$ for each successive (basal) plane of tetrahedrals. In this basis, the symmetrized rate matrix only has a single entry based on the escape rate from a tetrahedral and the tetrahedral-to-tetrahedral jump,
\be
\begin{split}
\omega &=
\begin{pmatrix}
  \frac{1}{\sqrt{2}}&  -\frac{1}{\sqrt{2}}\\
\end{pmatrix}
\cdot
\begin{pmatrix}
  -\transt{t}{t}-3\transt{t}{o}&  \transt{t}{t}\\
  \transt{t}{t}&  -\transt{t}{t}-3\transt{t}{o}\\
\end{pmatrix}
\cdot
\begin{pmatrix}
  \frac{1}{\sqrt{2}}\\
  -\frac{1}{\sqrt{2}}\\
\end{pmatrix}
\\
&= -(2\transt{t}{t} + 3\transt{t}{o})
\end{split}
\label{eqn:HCPSymmetrizedRate}
\ee
and the velocity vector at a tetrahedral is $\pm \left(3\transt{t}{o}\frac{c}{8} - \transt{t}{t}\frac{c}{4}\right)\probt{t}^{1/2}\zv$. After projecting into the site vector basis, and solving for the bias-correction vector, we have
\be
\bv{\text{t}}\ox\gv{\text{t}} = -\frac{c^2}{32}
\frac{\left(3\transt{t}{o} - 2\transt{t}{t}\right)^2}%
{3\transt{t}{o} + 2\transt{t}{t}}\probt{t}\zv\ox\zv
\label{eqn:HCPBiasCorrection}
\ee
and thus the diffusivity from \Eqn{HCPDiffusivityUncorrelated} is
\be
\begin{split}
\D = 
3a^2(\xv\ox\xv + \yv\ox\yv)\probt{o}\transt{o}{t} + \frac{c^2}{4}\zv\ox\zv \probt{o}\transt{o}{o}
+ \frac{c^2}{4}\zv\ox\zv \probt{t}\frac{3\transt{t}{o}\transt{t}{t}}{3\transt{t}{o} + 2\transt{t}{t}}
\end{split}
\label{eqn:HCPdiffusivity}
\ee

Computing the activation energy tensor is complicated due to the correlation terms. The uncorrelated contribution to inverse temperature derivative is
\be
\begin{split}
-\frac{d\D}{d\beta} &= 
\frac{c^2}{4}\zv\ox\zv \probt{o}\transt{o}{o} \left(\Etranst{o}{o}-\Eave\right)
+ 2\left(\frac32 a^2\xv\ox\xv + \frac32 a^2 \yv\ox\yv\right) \probt{o}\transt{o}{t} \left(\Etranst{o}{t}-\Eave\right)\\
& + 3\cdot2\left(\frac{c^2}{64}\zv\ox\zv\right) \probt{o}\transt{o}{t} \left(\Etranst{o}{t}-\Eave\right)
+ \frac{c^2}{16}\zv\ox\zv \probt{t}\transt{t}{t} \left(\Etranst{t}{t}-\Eave\right)\\
\end{split}
\label{eqn:HCPUncorrelatedDerivEnergy}
\ee
The change in the velocity vector for the tetrahedral site is
\be
\dbv{\text{t}} = \pm \left(\frac{3c}{8}\transt{t}{o}\left[\Etranst{o}{t} - \frac12(E_\text{t}+\Eave)\right] - \frac{c}{4}\transt{t}{t}\left[\Etranst{t}{t} - \frac12(E_\text{t}+\Eave)\right]\right)\probt{t}^{1/2}\zv
\label{eqn:HCPBiasVectorDerivEnergy}
\ee
and for the symmetric rate matrix
\be
\domega = -\left(2\transt{t}{t}\left[\Etranst{t}{t}-E_\text{t}\right] + 3\transt{t}{o}\left[\Etranst{o}{t}-E_\text{t}\right]\right).
\label{eqn:HCPSymmetrizedRateDerivEnergy}
\ee
After we combine all of the terms into \Eqn{DeltaDiffusivityEnergy}, we have an activation energy tensor
\be
\begin{split}
\Eact &= \left(\Etranst{o}{t} - \Eave\right)(\xv\ox\xv + \yv\ox\yv) + \\
& \left(\probt{o}\transt{o}{o}+ \probt{t}\frac{3\transt{t}{o}\transt{t}{t}}{3\transt{t}{o} + 2\transt{t}{t}}\right)^{-1}\Bigg[ \left\{\Etranst{o}{o} - \Eave\right\}\probt{o}\transt{o}{o}\\
&+ \left\{\frac{3\transt{t}{o}}{3\transt{t}{o}+2\transt{t}{t}}\left(\Etranst{t}{t}-\Eave\right)
+ \frac{2\transt{t}{t}}{3\transt{t}{o}+2\transt{t}{t}}\left(\Etranst{t}{o}-\Eave\right)\right\} \probt{t}\frac{3\transt{t}{o}\transt{t}{t}}{3\transt{t}{o} + 2\transt{t}{t}} \Bigg](\zv\ox\zv).
\end{split}
\label{eqn:HCPActivationEnergy}
\ee
For basal plane diffusivity, the activation barrier is $(\Etranst{o}{t}-\Eave)$, as expected. For diffusion along the $c$-axis, the expression in \Eqn{HCPActivationEnergy} simplifies in several key limits. If $\probt{o}\transt{o}{o}$ is the fastest, then $\Eact \approx (\Etranst{o}{o}-\Eave)$; diffusion is dominated by direct octahedral-to-octahedral jumps. If it is not, then the octahedral-tetrahedral network dominates; if $\transt{t}{t}$ is rate-limiting (slower than $\transt{t}{o}$), $\Eact \approx (\Etranst{o}{t}-\Eave)$, otherwise $\Eact \approx (\Etranst{t}{t} - \Eave)$. These important limits are all captured in a single expression for the activation barrier, and can be computed automatically.

The elastic dipole terms break into basal and $c$-axis components, with a tetragonal component for the octahedral-tetrahedral jump. The elastic dipole associated with this octahedral has two components, due to hexagonal symmetry,
\beu
\Ps{\text{o}} = 
\Pst{o}{b}(\xv\ox\xv+\yv\ox\yv) + \Psc{o}{c}\zv\ox\zv,
\label{eqn:HCPoctelasticdipole}
\eeu
corresponding to the basal $\Pst{o}{b}$ and $c$-axis $\Psc{o}{c}$ components. The octahedral-to-octahedral transition along the $c$-axis behaves similarly,
\beu
\PTt{o}{o} = 
\PTc{o}{o}{\text{b}}(\xv\ox\xv + \yv\ox\yv) + \PTc{o}{o}{c}\zv\ox\zv.
\label{eqn:HCPooelasticdipole}
\eeu
The elastic dipole associated with the tetrahedral has two components,
\beu
\Ps{\text{t}} = 
\Pst{t}{b}(\xv\ox\xv+\yv\ox\yv) + \Psc{t}{c}\zv\ox\zv,
\label{eqn:HCPtetelasticdipole}
\eeu
corresponding to the basal $\Pst{t}{b}$ and $c$-axis $\Psc{t}{c}$ components. The tetrahedral-to-tetrahedral transition along the $c$-axis also has hexagonal symmetry,
\beu
\PTt{t}{t} = 
\PTc{t}{t}{\text{b}}(\xv\ox\xv + \yv\ox\yv) + \PTc{t}{t}{c}\zv\ox\zv.
\label{eqn:HCPttelasticdipole}
\eeu
Finally, the octahedral-to-tetrahedral transition state does not locally have hexagonal symmetry. If we consider one jump along $\frac{a}{3}[1\bar10\frac{3c}{8a}]$, oriented so that it lies in the $yz$ plane, then we can write the three components of the elastic dipole,
\beu
\PTt{o}{t} = 
\PTc{o}{t}{\text{b}}(\xv\ox\xv + \yv\ox\yv) + \PTc{o}{t}{\text{t}}(-\xv\ox\xv+\yv\ox\yv) + \PTc{o}{t}{c}\zv\ox\zv,
\label{eqn:HCPotelasticdipole}
\eeu
corresponding to a basal, tetragonal, and $c$-axis contributions. Alternatively, we could write contributions as $\PTc{o}{t}{\|} = (\PTc{o}{t}{\text{b}}+\PTc{o}{t}{\text{t}})\yv\ox\yv$ along the hop and $\PTc{o}{t}{\perp} = (\PTc{o}{t}{\text{b}}-\PTc{o}{t}{\text{t}})\xv\ox\xv$ perpendicular to the hop. We find that, after applying the 3-fold rotation axis that the tetragonal component only appears in our bare term.

Hexagonal symmetry reduces the elastodiffusion tensor to seven independent components, of which two have contributions from correlation. The contributions from correlation come from the tetrahedral site. The change in the velocity vector for the tetrahedral site is
\begin{multline}
  \dbv{\text{t}} = \pm \left(\frac{3c}{8}\transt{t}{o}\left[\PTc{o}{t}{\text{b}} - \frac12(\Pst{t}{b}+\Pave_\text{b})\right] - \frac{c}{4}\transt{t}{t}\left[\PTc{t}{t}{\text{b}} - \frac12(\Pst{t}{b}+\Pave_\text{b})\right]\right)\probt{t}^{1/2}(\zv\ox\xv\ox\xv + \zv\ox\yv\ox\yv) \\
    \pm \left(\frac{3c}{8}\transt{t}{o}\left[\PTc{o}{t}{c} - \frac12(\Psc{t}{c}+\Pave_c)\right] - \frac{c}{4}\transt{t}{t}\left[\PTc{t}{t}{c} - \frac12(\Psc{t}{c}+\Pave_c)\right]\right)\probt{t}^{1/2}\zv\ox\zv\ox\zv
\label{eqn:HCPBiasVectorDerivStrain}
\end{multline}
and for the symmetric rate matrix
\begin{multline}
\domega = -\left(2\transt{t}{t}\left[\PTc{t}{t}{\text{b}}-\Pst{t}{b}\right] + 3\transt{t}{o}\left[\PTc{t}{t}{\text{b}}-\Pst{t}{b}\right]\right)(\xv\ox\xv+\yv\ox\yv)\\
-\left(2\transt{t}{t}\left[\PTc{t}{t}{c}-\Psc{t}{c}\right] + 3\transt{t}{o}\left[\PTc{t}{t}{c}-\Psc{t}{c}\right]\right)\zv\ox\zv,
\label{eqn:HCPSymmetrizedRateDerivStrain}
\end{multline}
where $\Pave = \Ps{\text{o}}\probt{o} + \Ps{\text{t}}2\probt{t}$, as in our FCC lattice. Symmetry gives us 6 unique elastodiffusion components, as basal-plane isotropy requires $2\ddc{66}=\ddc{11}-\ddc{22}$. Moreover, only the $\ddc{31}=\ddc{32}$ and $\ddc{33}$ components have contributions from correlations; only $\ddc{11}=\ddc{22}$, $\ddc{33}$, $\ddc{44}=\ddc{55}$ and $\ddc{66}$ have contributions from the geometric terms; and $\ddc{44}=\ddc{55}$ \textit{only} has contributions from geometry. Using the Voigt notation for fourth-rank tensors with \Eqn{DeltaElastoDiffusion},
\be
\begin{split}
  \ddc{11} = \ddc{22} &= a^2\left\{ 6 + \beta\left(\PTc{o}{t}{\text{b}} + \frac12 \PTc{o}{t}{\text{t}}  - \Pave_\text{b}\right)\right\}\probt{o}\transt{o}{t}\\
  \ddc{33} &= \frac{c^2}{4}\Bigg\{ \left[2 + \beta\left(\PTc{o}{o}{\text{c}} - \Pave_\text{c}\right)\right]\probt{o}\transt{o}{o} \\
  &+ \left[2 + \frac{3\transt{t}{o}}{3\transt{t}{o}+2\transt{t}{t}}\left(\PTc{t}{t}{\text{c}}-\Pave_\text{c}\right)
+ \frac{2\transt{t}{t}}{3\transt{t}{o}+2\transt{t}{t}}\left(\PTc{t}{o}{\text{c}}-\Pave_\text{c}\right)\right] \probt{t}\frac{3\transt{t}{o}\transt{t}{t}}{3\transt{t}{o} + 2\transt{t}{t}} \Bigg\}\\
  \ddc{12} = \ddc{21} &= a^2\left\{ \beta\left(\PTc{o}{t}{\text{b}} - \frac12 \PTc{o}{t}{\text{t}}  - \Pave_\text{b}\right)\right\}\probt{o}\transt{o}{t}\\
  \ddc{13} = \ddc{23} &= a^2\left\{ \beta\left(\PTc{o}{t}{\text{c}} - \Pave_\text{c}\right)\right\}\probt{o}\transt{o}{t}\\
  \ddc{31} = \ddc{32} &= \frac{c^2}{4}\Bigg\{ \beta\left(\PTc{o}{o}{\text{b}} - \Pave_\text{b}\right)\probt{o}\transt{o}{o} \\
  &+ \left[ \frac{3\transt{t}{o}}{3\transt{t}{o}+2\transt{t}{t}}\left(\PTc{t}{t}{\text{b}}-\Pave_\text{b}\right)
    + \frac{2\transt{t}{t}}{3\transt{t}{o}+2\transt{t}{t}}\left(\PTc{t}{o}{\text{b}}-\Pave_\text{b}\right)
    \right]\probt{t}\frac{3\transt{t}{o}\transt{t}{t}}{3\transt{t}{o} + 2\transt{t}{t}} \Bigg\}\\
  \ddc{44} = \ddc{55} &=
\frac{3a^2}{2}\probt{o}\transt{o}{t} + \frac{c^2}{8}\probt{o}\transt{o}{o}
+ \frac{c^2}{8}\probt{t}\frac{3\transt{t}{o}\transt{t}{t}}{3\transt{t}{o} + 2\transt{t}{t}}\\
  \ddc{66} &= \frac{\ddc{11}-\ddc{12}}{2} = a^2\left\{ 3 + \frac\beta2\PTc{o}{t}{\text{t}}\right\}\probt{o}\transt{o}{t}\\
\end{split}
\label{eqn:HCPelastodiffusion}
\ee
and all other terms are zero. The activation volume tensor can be derived as before; the scalar activation volume from \Eqn{ScalarActivationVolume} is
\be
\begin{split}
  V^\text{act} =& \frac{\kB T}{9B}\Bigg\{
\left(D_\text{b}^{-1}\ddc{11} + D_\text{b}^{-1}\ddc{12}+D_\text{c}^{-1}\ddc{31}\right)
\frac{2S_{11}+2S_{12}+2S_{13}}{2S_{11}+2S_{12}+4S_{13}+S_{33}}\\
&\quad+ \left(2D_\text{b}^{-1}\ddc{13} + D_\text{c}^{-1}\ddc{33}\right)
\frac{2S_{13}+S_{33}}{2S_{11}+2S_{12}+4S_{13}+S_{33}}
\Bigg\},
\end{split}
\label{eqn:HCPscalarActivationVolume}
\ee
where the bulk modulus $B^{-1} = 2S_{11}+2S_{12}+4S_{13}+S_{33}$.

\subsection{Carbon in iron}
Carbon in body-centered cubic iron sits on octahedral sites, and transitions through tetrahedral transition states. Veiga~\et\cite{Veiga2011} used an EAM potential to compute the octahedral and tetrahedral energies and elastic dipoles. The lattice constant for this potential is $a_0 = 2.8553\text{ \AA}$, and the elastic constants are $C_{11} = 243\text{ GPa}$, $C_{12} = 145\text{ GPa}$, and $C_{44} = 116\text{ GPa}$. The tetrahedral transition state is 0.816 eV above the octahedral site, and the attempt frequency is taken as 10 THz\cite{Veiga2011}. The dipole tensors can be separated into parallel and perpendicular components; the parallel direction points towards the closest Fe atoms for the C, while the perpendicular components lie in the interstitial plane. For the octahedral, the parallel component is 8.03 eV, and the perpendicular components are both 3.40 eV; for the tetrahedral transition state, the parallel component is 4.87 eV, and the perpendicular are both 6.66 eV.

\begin{figure}[ht]
  \begin{center}
  \includegraphics[width=8cm]{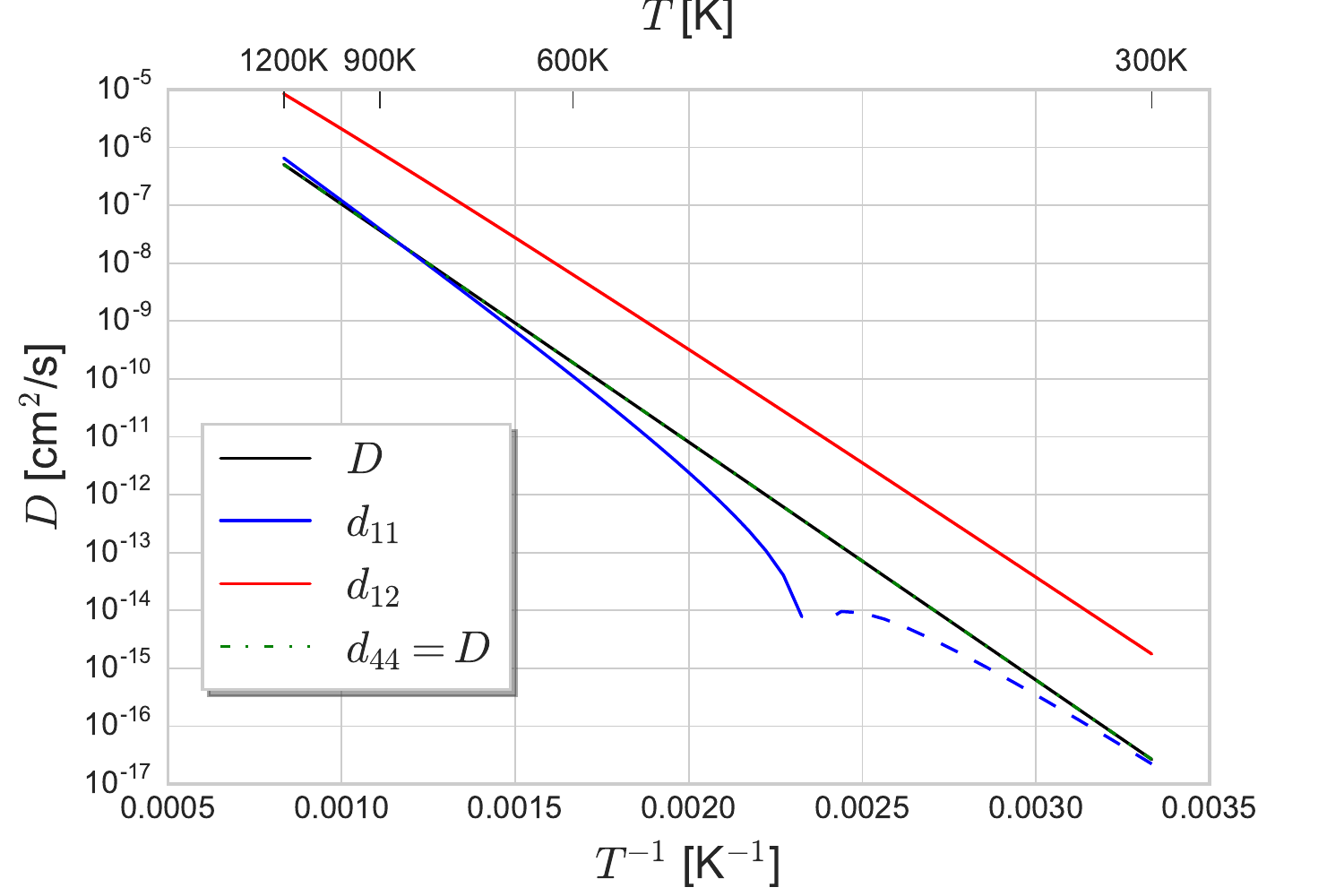}
  \end{center}
\caption{Diffusivity and elastodiffusivity of C in BCC Fe between 300K and 1200K. The carbon diffusivity is Arrhenius, with $D(T) = 1.359\EE{-3}\exp(-0.816/\kB T)\text{ cm}^2/\text{s}$. The derivative $d_{11} = dD_{xx}/d\eps_{xx}$ changes sign: negative below $\approx 425\text{K}$, and positive above. The derivative $d_{44} = dD_{xy}/d\eps_{xy} = D$, as it only has a contribution from the geometric change with strain.}
\label{fig:C-diffusivity}
\end{figure}

\begin{figure}[ht]
  \begin{center}
  \includegraphics[width=8cm]{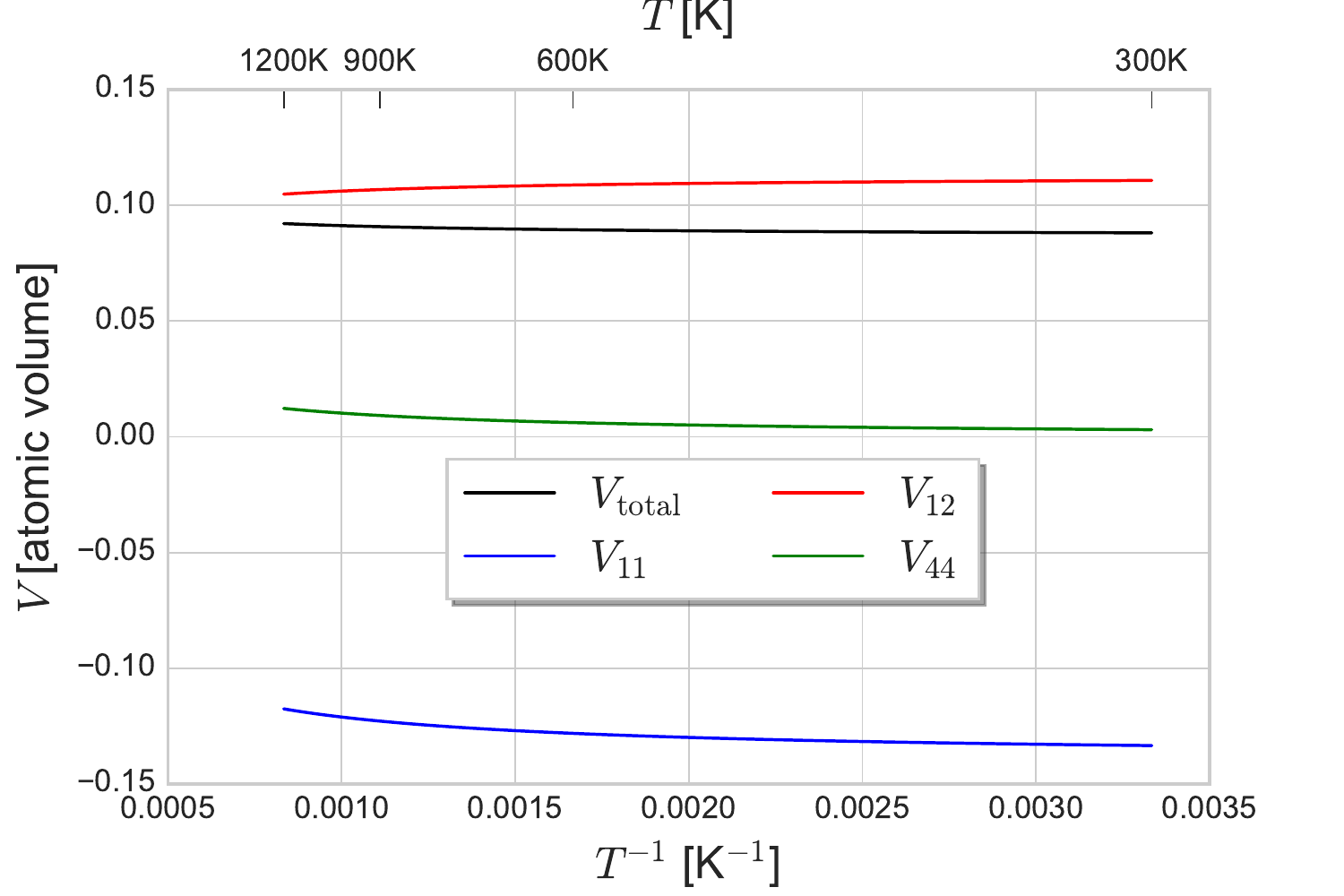}
  \end{center}
  \caption{Activation volume of C in BCC Fe. The total activation volume from \Eqn{ScalarActivationVolume} is the sensitivity to the isotropic diffusivity to hydrostatic pressure; at 300K it is $0.0307\Omega_\text{Fe} = 0.357\text{ \AA}^3$. The components from \Eqn{ActivationVolume} have cubic symmetry, and so the representative terms at 300K are $V_{11} = V_{xxxx} = -0.1175\Omega_\text{Fe} = -1.368\text{ \AA}^3$, $V_{12} = V_{xxyy} = 0.1048\Omega_\text{Fe} = 1.220\text{ \AA}^3$, and $V_{44} = 2V_{xyxy} = 0.0122\Omega_\text{Fe} = 0.142\text{ \AA}^3$. The only contribution to $V_{xyxy}$ is geometric.}
\label{fig:C-activation-volume}
\end{figure}

\Fig{C-diffusivity} shows the Arrhenius behavior of the diffusivity, while \Fig{C-activation-volume} reveals unusual behavior in the activation volume from 300K to 1200K. The diffusivity follows a simple Arrhenius relationship. The elastodiffusion tensor has three unique components: $d_{11}=dD_{xx}/d\eps_{xx}$, $d_{12}=dD_{xx}/d\eps_{yy}$, $d_{44}=dD_{xy}/d\eps_{xy}$. The $d_{44}$ term only contains geometric contributions, while $d_{11}$ changes sign as the (positive) geometric contribution dominates at high temperatures and the (negative) activation contribution dominates at low temperatures. The $d_{12}$ term is an order of magnitude larger than the other elastodiffusivity coefficients, and also leads to the unusual activation volume behavior where $V_{11}$ is negative, $V_{12}$ is positive, and $V_\text{total} = V_{11}+2V_{12}$ is positive. This means that for a compressive uniaxial stress along $\langle100\rangle$, the diffusivity for carbon \textit{increases} along $\langle100\rangle$, while the perpendicular diffusivities \textit{decrease}. The unusual behavior is caused by the Poisson effect combined with the large $d_{12}$ value and small $d_{11}$ value. This still occurs while a hydrostatic compressive stress decreases the diffusivity.

\section{Conclusions}
The general expression for the elastodiffusion tensor, activation energy and volume tensors build on previous work that assumed interstitial migration without correlation. As shown here, correlation contributions must be included even in cases of interstitial diffusion where dictated by the diffusion network. The application of this method extends beyond interstitial solutes in a crystal, but can also be used for more complex cases such as self-interstitials (which may possess multiple intermediate states) or diffusion through boundaries. There is an open-source numerical implementation of the diffusion, elastodiffusion, and activation barrier tensors, described in Appendix~\ref{sec:implementation}\cite{OnsagerCalc}. The perturbation analysis laid out here can also be applied to error propagation, by calculating the derivative of diffusivity with respect to site and transition energies. It should be possible, though more complicated, to derive similar expressions for the elastodiffusion tensor for more complex diffusion mechanisms, such as vacancy-mediated solute diffusion. 

\section*{Acknowledgements}
The author thanks Maylise Nastar and Pascal Bellon for helpful conversations.

\section*{Disclosure}
The author has no conflicts of interest.

\section*{Funding}
Research supported in part by the U.S. Department of Energy, Office of Basic Energy Sciences, Division of Materials Sciences and Engineering under Award \#DE-FG02-05ER46217, through the Frederick Seitz Materials Research Laboratory, in part by the Office of Naval Research grant N000141210752, and in part by the National Science Foundation Award 1411106. Part of this research was performed while the author was visiting the Institute for Pure and Applied Mathematics (IPAM) at UCLA, which is supported by the National Science Foundation (NSF).


\appendices
\section{Pseudoinverse perturbation}
\label{app:PseudoInverse}
The expression in \Eqn{DeltaGreenEnergy}, which is similarly used to express the first-order change in pseudoinverse of $\omega$ with respect to strain, comes from the perturbation of the pseudoinverse and lacks terms that would correspond to the perturbative change in the null space. In this case, the simplified expression is due to the symmetry of the matrix $\omega$, the perturbative changes $\delta\omega$, and the symmetry of the vector dot-products with $\bv{}$. If a general matrix $A$ has pseudoinverse $A^+$, then we can expand the pseudoinverse $B^+ = (A+E)^+$ up to first order in the perturbation $E$ as
\be
B^+ = A^+ - A^+(AA^+)E(A^+A)A^+ + (A^\dagger A)^+(A^+A)E^\dagger(\mathbf{1} - AA^+) - (\mathbf{1} - A^+A)E^\dagger (AA^+)(AA^\dagger)^+ + O(\|E\|^2)
\label{eqn:GeneralizedPseudoInverse}
\ee
from Eqn. (3.24) of \cite{Stewart1977}. For our case of interest, $A=\omega$ and $A^+ = g$ are both real symmetric matrices; moreover, we want to compute $\vv\cdot B^+\cdot\vv$ for an arbitrary vector $\vv$,
\be
\begin{split}
\vv\cdot B^+\cdot\vv &= \vv\cdot A^+\cdot\vv - \vv\cdot(A^+ E A^+)\cdot\vv\\
& + \vv\cdot\left\{(A^2)^+(A^+A)E(\mathbf{1} - AA^+) - \left[(A^2)^+(A^+A)E(\mathbf{1} - AA^+)\right]^\text{T}\right\}\cdot\vv + O(\|\vv\cdot E\}^2)\\
 &= \vv\cdot A^+\cdot\vv - \vv\cdot(A^+ E A^+)\cdot\vv + O(\|\vv\cdot E\}^2)\\
\end{split}
\label{eqn:SymmetricPseudoInverse}
\ee
which, for our problem, reduces to first order in $\delta\omega$
\be
\bv{}\cdot \delta g\cdot\bv{} = -\bv{} \cdot g\,\delta\omega\,g\cdot\bv{} = - \gv{}\cdot\delta\omega\cdot\gv{}.
\label{eqn:DiffusionPseudoInverse}
\ee

\section{Vector and tensor bases at sites}
To construct the vector and tensor bases, we first must determine the eigenvectors of each group operation. There are only ten types of operations available in a crystal: identity, a 2-, 3-, 4-, or 6-fold axis, mirror through a plane, or a 2-, 3-, 4-, or 6-fold axis combined with a mirror through that same plane. These can be identified from the determinant---which is $-1$ if the operation contains a mirror operation or 1 if it does not---and the trace,
\be
\begin{matrix}
\det&\Tr&\text{operation}&&\det&\Tr&\text{operation}\\
\hline
1&3&\text{identity}&&-1&1&\text{mirror}\\
1&2&\text{6-fold axis}&&-1&0&\text{6-fold axis + mirror}\\
1&1&\text{4-fold axis}&&-1&-1&\text{4-fold axis + mirror}\\
1&0&\text{3-fold axis}&&-1&-2&\text{3-fold axis + mirror}\\
1&-1&\text{2-fold axis}&&-1&-3&\text{inversion}\\
\hline\\
\end{matrix}
\label{eqn:SymmOperationTypes}
\ee
The eigenvalues and eigenvectors for these cases are all straightforward. The trivial cases are identity---triply-degenerate eigenvalues of 1, for the three eigenvectors $\ev_1$, $\ev_2$, $\ev_3$---inversion---triply-degenerate eigenvalues of $-1$, for the three eigenvectors  $\ev_1$, $\ev_2$, $\ev_3$---and mirror---one eigenvalue of $-1$ with eigenvector $\vv$ corresponding to the mirror plane normal and double-generate eigenvalues of 1 for the two orthonormal vectors $\vv'$ and $\vv''$ that are orthogonal to $\vv$. For the remaining cases, the rotational axis $\vv$ is always an eigenvector with an eigenvalue $+1$ if there is not a mirror operation, or $-1$ if there is a mirror operation. The remaining eigenvalues are two $n^\text{th}$ roots of unity $e^{\pm i2\pi/n}$ for the $n$-fold rotation axis. If $\vv'$ and $\vv''$ are two orthonormal vectors that are also orthogonal to the rotational axis $\vv$, then the eigenvectors are $(\vv' \pm i\vv'')/\sqrt{2}$ for the positive and negative roots of unity.

We construct vector and tensor bases that are compatible with a set of point group operations from the eigenvectors of the group operations. The bases are such that they remained unchanged under application of every group operation. For each group operation, $d$-rank tensors can be constructed from different dyad products of eigenvectors of the group operation. In particular, the \textit{product} of the eigenvalues of the eigenvectors needs to be 1. Hence, for a vector basis (a 1-rank tensor), we have: (a) identity produces a 3-dimensional basis of $\{\ev_1, \ev_2, \ev_3\}$, (b) mirror produces a 2-dimensional basis of $\{\vv', \vv''\}$, two orthonormal vectors in the mirror plane, (c) $n$-fold rotational axes without mirror produces a 1-dimensional basis of $\{\vv\}$, the rotational axis vector, and (d) all other cases produce a 0-dimensional null basis, $\varnothing$. Each group operation produces its own basis, and these vector bases are then intersected to produce the final basis, using some simple rules: if $p$- and $q$-dimensional bases are intersected, the resulting dimension of the final basis is not greater than $\min\{p, q\}$. Then, the remaining non-trivial cases are: two 2-dimensional, a 2- and 1-dimensional, and two 1-dimensional intersections. In the first case, if $\vv$ and $\vv'$ are the vectors made by the cross-product of the two vectors in each basis, then the resulting intersection will be either (a) the same 2-dimensional basis if $\vv$ is parallel to $\vv'$, or (b) a 1-dimensional basis consisting of $\vv\times\vv'$, otherwise. In the second case, if $\vv_1$ is the basis for the 1-dimensional basis, and $\vv_2$ is the cross-product of the two vectors in the 2-dimensional basis, then the resulting intersection will be either (a) 1-dimensional basis consisting of $\vv_1$ if $\vv_1$ is perpendicular to $\vv_2$, or (b) the 0-dimensional empty basis $\varnothing$, otherwise. Finally, if $\vv$ and $\vv'$ are the two 1-dimensional bases, the resulting intersection will either be (a) $\vv$ if $\vv$ is parallel to $\vv'$, or (b) the 0-dimensional empty basis $\varnothing$, otherwise.

The construction of the second-rank tensor bases is more complicated, but reduces to three distinct cases. We are specifically interested in \textit{symmetric} second-rank tensors, so the largest the bases can be is 6-dimensional. For identity and inversion, the full 6-dimensional basis is invariant: $\{\ev_1\ox\ev_1, \ev_2\ox\ev_2, \ev_3\ox\ev_3, (\ev_1\ox\ev_2 + \ev_2\ox\ev_1)/\sqrt{2}, (\ev_1\ox\ev_3 + \ev_3\ox\ev_1)/\sqrt{2}, (\ev_2\ox\ev_3 + \ev_3\ox\ev_2)/\sqrt{2}\}$. For a mirror operation or a 2-fold rotation, if $\vv$ is the mirror plane normal / rotational axis, and $\vv'$ and $\vv''$ are the remaining orthonormal vectors, the 4-dimensional basis is invariant: $\{\vv\ox\vv, \vv'\ox\vv', \vv''\ox\vv'', (\vv'\ox\vv'' + \vv''\ox\vv')/\sqrt{2}\}$. Finally, for the $n$-fold rotations (with or without a mirror), if $\vv$ is the rotational axis, and $\vv'$ and $\vv''$ are the remaining orthonormal vectors, the 2-dimensional basis is invariant: $\{\vv\ox\vv, (\vv'\ox\vv' + \vv''\ox\vv'')/\sqrt{2}\}$. To intersect any two bases, the simplest approach is to find the null-space of the combined column spaces constructed from the bases using singular-value decomposition\cite{JohnsonLinearAlgebra2002}.

\section{Implementation}
\label{sec:implementation}
A full numerical implementation of the algorithms in Python described are available on github\cite{OnsagerCalc} under the MIT License. This includes algorithms to analyze a given crystal (lattice and atomic basis), find generators for the space group operations, determine all point group operations for each site, identify Wyckoff positions, generate vector and tensor bases, construct a jump network in the crystal, and identify unique jumps. Once the energies, prefactors, and elastic dipoles are determined for the unique sites and jumps, the numerical implementation can compute the diffusivity, elastodiffusion, and activation barrier tensors for a given temperature. The implementation includes numeric examples of the analytic results from this paper, with input files for FCC, BCC, and HCP interstitial diffusion.

\end{document}